\documentclass{elsarticle}
\usepackage{graphicx}
\usepackage[intlimits]{amsmath}
\usepackage{amssymb,amsfonts}
\usepackage{bbm}
\usepackage{booktabs}
\usepackage{mathrsfs}
\usepackage{textcomp}
\usepackage[T1]{fontenc}
\usepackage{lmodern}
\usepackage[nottoc]{tocbibind}

\renewcommand{\vec}[1]{\boldsymbol{#1}}

\newcommand{\im}{\text{i}}

\newcommand{\La}{\ensuremath{\mathcal L}}

\def\be{\begin{equation}}
\def\ee{\end{equation}}
\def\bg{\begin{eqnarray}}
\def\en{\end{eqnarray}}
\def\nn{\nonumber}

\journal{Nuclear Physics A}

\begin{document}
\begin{frontmatter}

\title{Production of cascade hypernuclei via the ($K^-,K^+$) reaction 
within a quark-meson coupling model} 
\author[CSSM,Saha]{R. Shyam} 
\author[CSSM]{K. Tsushima} 
\author[CSSM]{A.W. Thomas}
\address[CSSM]{ Centre for the Subatomic Structure of Matter (CSSM),
School of Chemistry and Physics, University of Adelaide, SA 5005, Australia}
\address[Saha]{Saha Institute of Nuclear Physics, 1/AF Bidhan Nagar, 
Kolkata 700064, India} 


\begin{abstract}
We study the production of bound hypernuclei $^{12}{\!\!\!_{\Xi^-}}$Be and 
$^{28}{\!\!\!_{\Xi^-}}$Mg via the $(K^-,K^+)$ reaction on $^{12}$C and
$^{28}$Si targets, respectively, within a covariant effective Lagrangian 
model, employing $\Xi$ bound state spinors derived from the latest 
quark-meson coupling model as well as Dirac single particle wave 
functions. The $K^+\Xi^-$ production vertex is described by excitation, 
propagation and decay of $\Lambda$ and $\Sigma$ resonance states in the 
initial collision of a $K^-$ meson with a target proton in the incident 
channel. The parameters of the resonance vertices are fixed by describing 
the available data on total and differential cross sections for the 
$p(K^-, K^+)\Xi^-$ reaction. We find that both the elementary and 
hypernuclear production cross sections are dominated by the contributions
from the $\Lambda$(1520) intermediate resonant state. The $0^\circ$ 
differential cross sections for the formation of simple s-state $\Xi^-$ 
particle-hole states peak at a beam momentum around 1.0 GeV/c, with a 
value in excess of 1 $\mu$b.   
\end{abstract}

\end{frontmatter}
\section{Introduction}

The study of the double strangeness ($S$) hypernuclei is of decisive 
importance for revealing the entire picture of strong interactions among
octet baryons. The binding energies and widths of the $\Xi$ hypernuclear 
states are expected to determine the strength of the $\Xi N$ and $\Xi N \to 
\Lambda \Lambda$ interactions, respectively. This basic information is key 
to testing the quark exchange aspect of the strong interaction because long 
range pion exchange plays essentially a very minor role in the $S = -2$ 
sector. Even though t-channel pion exchange between $\Xi$ and nucleon does
operate, its strength is quite weak because the $\pi \Xi \Xi$ coupling is 
smaller as compared to the $\pi NN$ coupling~\cite{rij10}. This input is also 
vital for understanding the multi-strange hadronic or quark matter. Since 
strange quarks are negatively charged they are preferred in charge neutral dense 
matter. Thus these studies are of crucial value for investigating the role 
of strangeness in the equation of state at high density, as probed in the 
cores of neutron stars~\cite{bie10,lat04} and in high energy heavy ion 
collisions at relativistic heavy ion colliders (RHIC) at Brookhaven National 
laboratory~\cite{ada07}, CERN~\cite{aam11} and FAIR facility at 
GSI~\cite{CBM11}.

The $(K^-,K^+)$ reaction leads to the transfer of two units of both charge 
and strangeness to the target nucleus. Thus this reaction is one of 
the most promising ways of studying the $S = -2$ systems such as $\Xi$ 
hypernuclei and a dibaryonic resonance ($H$), which is a near stable 
six-quark state with spin parity of $0^+$ and isospin 0
\cite{jaf77,mul83,bea11}. Several ways have been discussed to approach 
these systems in the past~\cite{aer82,aer83}. Many experimental groups 
have used the $(K^-,K^+)$ reaction on nuclear targets to search for a $H$
dibaryonic resonance~\cite{iim92,fuk98,ahn98,kha00,yoo07}. 

As far as $\Xi$ hypernuclei are concerned, there are some hints of their 
existence from emulsion events~\cite{wil59}. However,  no $\Xi$ bound 
state was unambiguously observed in the few experiments performed involving 
the ($K^-,K^+)$ reaction on a $^{12}$C target~\cite{fuk98,kha00} because of
the limited statistics and detector resolution. However, in the near 
future experiments will be performed at the JPARC facility in Japan to
observe the bound states of $\Xi$ hypernuclei via the $(K^-,K^+)$ reaction
with the best energy resolution of a few MeV and with large statistics by
using the newly constructed high-resolution spectrometers~\cite{nag06}.
The first series of experiments will be performed on a $^{12}$C target.
These measurements are of great significance because convincing evidence
for the $\Xi$ single-particle bound states would yield vital information
on $\Xi$ single particle potential and the effective $\Xi N$ interaction.
Already, the analysis of the scarce emulsion~\cite{gal83} and spectrometer 
data~\cite{fuk98,kha00} have led to $\Xi$-nuclear potentials with depths 
that differ by about 10 MeV from each other.

The $(K^-,K^+)$ reaction implants a $\Xi$ hyperon in the nucleus through 
the elementary process $p(K^-, K^+)\Xi^-$. The cross sections for 
the elementary reaction were measured in the 1960s and early 1970s using 
hydrogen bubble chambers~\cite{pje62,ber66,mer68,bur68,dau69,car73}. The 
total cross-section data from these measurements are tabulated in 
Ref.~\cite{fla83}. In a recent study~\cite{shy11}, this  reaction was 
investigated within a single-channel effective Lagrangian model where 
contributions were included from the $s$-channel [see, Fig. 1(a)] and 
$u$-channel diagrams which have as intermediate states $\Lambda$ and 
$\Sigma$ hyperons together with eight of their three-and four-star 
resonances with masses up to 2.0 GeV [$\Lambda(1405)$, $\Lambda(1520)$, 
$\Lambda(1670)$, $\Lambda(1810)$, $\Lambda(1890)$, $\Sigma(1385)$, 
$\Sigma(1670)$ and $\Sigma(1750)$, which are represented by $\Lambda^*$ 
and $\Sigma^*$ in Fig.~1a]. This reaction is a clean example of a process 
in which baryon exchange plays the dominant role and the $t$-channel meson 
exchanges are absent, as no meson with $S=+2$ is known to exist. An important 
observation of that study is that the total cross section of the 
$p(K^-, K^+)\Xi^-$ reaction is dominated by the contributions from the 
$\Lambda(1520)$ (with $L_{I 2J}=D_{03}$) resonance intermediate state 
through both $s$- and $u$-channel terms. The region for beam momentum 
($p_{K^-}$) below 2.0 GeV/c was shown to get most contributions from the 
$s$-channel graphs - the $u$-channel terms are dominant only in the region 
$p_{K^-}$ > 2.5 GeV. 
\begin{figure}[t]
\centering
\includegraphics[width=.80\textwidth]{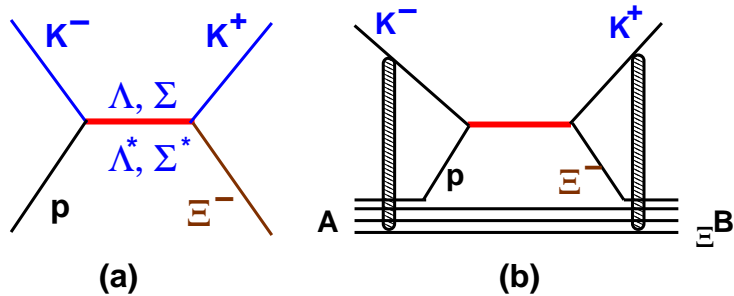}
\caption{(color online)
Graphical representation of our model to describe $p(K^-, K^+)\Xi^-$ 
(Fig.~1a) and $A(K^-, K^+){_{\Xi^-}}$\!\! B reactions (Fig. 2b). In the
latter case the shaded area depicts the optical model interactions in the
incoming and outgoing channels.   
}
\label{fig:Fig1}
\end{figure}

Almost all of the previous theoretical investigations of cascade 
hypernuclear production via $(K^-,K^+)$ reaction on target nuclei 
\cite{gal83,yam94,ike94,tad95,har10} have used the framework of an impulse 
approximation where the hyperon production dynamics is separated from that 
of the relative motion in the entrance and outgoing channels. Thus, the 
hypernuclear production cross section is expressed as a product of the 
cross section of the elementary cascade production reaction and a term that
accounts for the dynamics of the relative motion. None of these models has 
attempted to calculate the cross sections of the elementary reaction -  
they have been extracted from the sparse experimental data. Therefore, the 
results of these calculations carry over the ambiguities that are involved
in the experimental values of the differential cross sections for the 
elementary reactions. 
     
In this paper,  we investigate the production of cascade hypernuclei via 
the $(K^-,K^+)$ reaction on nuclear targets within an effective Lagrangian 
model~\cite{shy04,shy08,shy09}, which is similar to that used in Ref.
\cite{shy11} to study the elementary production reaction, 
$p(K^-, K^+)\Xi^-$. We consider only the $s$-channel production diagrams 
(see Fig.~1b) as we are interested in the region where $p_{K^-}$ lies 
below 2 GeV/c. The model retains the full field theoretic structure of the 
interaction vertices and treats baryons as Dirac particles. The initial 
state interaction of the incoming $K^-$ with a bound target proton leads to 
excitation of intermediate $\Lambda$ and $\Sigma$ resonant states, which 
propagate and subsequently decay into a $\Xi^-$ hyperon that gets captured
into one of the nuclear orbits, while the other decay product, the $K^+$ 
goes out. In Ref.~\cite{shy11}, it was shown that six intermediate resonant 
states, $\Lambda$, $\Lambda(1405)$, $\Lambda(1520)$, $\Lambda(1810)$, 
$\Sigma$, and $\Sigma(1385)$,  make the most significant contributions to 
the cross sections of the elementary process. Therefore, in our present 
study the amplitudes corresponding to these six resonant states have been 
considered.

\section{Formalism}
\subsection{bound state spinors}

The $\Xi^-$ bound state spinors have been calculated in the quark-meson 
coupling (QMC) model as well as in a phenomenological model where they are 
obtained by solving the Dirac equation with scalar and vector fields having 
a Woods-Saxon (WS) radial form. In the latter case, with a set of radius and 
diffuseness parameters, the depths of these fields are searched to reproduce 
the binding energy (BE) of a given state. Since the experimental values of 
the BEs for the $\Xi^-$ bound states are as yet unknown, we have adopted the 
corresponding QMC model predictions (as shown in table 1) in our search 
procedure for these states. Furthermore, the scalar and vector fields are 
assumed to have the same geometry. It should be noted that the depths of the 
potential fields in such a model are dependent on the adopted radius ($r_0$) 
and diffuseness ($a$) parameters but there is no certain way of fixing them. 
Nevertheless, using same $r_0$ for all the states may make the search for the 
potential depths too restrictive. Some authors have used the root mean square 
radius (RMS) of a given state to fix the $r_0$ parameter (see, e.g., Refs.
\cite{pet98} and~\cite{ben10}). However, such a procedure cannot be applied 
for the $\Xi$ bound states at this stage due to the lack of any experimental 
information about them. With these constraints, we show in Table 1 the 
resulting parameters associated with the scalar and vector fields of the 
phenomenological model for $\Xi^-$ bound and proton hole states for the two 
target nuclei. In this table the QMC predictions for the BE of the proton 
hole states are also shown. However, in the search procedure for these states 
the experimental values of the BEs (given within the brackets) have been used. 
\begin{table}[t]
\centering
\caption{\label{table1} Parameters of the Dirac single particle potential 
(having a WS radial shape) for the $\Xi^-$ bound and proton hole states. In 
each case radius ($r_0$) and diffuseness ($a$) parameters were 0.983 fm, 
and 0.606 fm, respectively for both vector and scalar potentials. The binding
energies (BEs) of the $\Xi^-$ states were taken from the predictions of the 
QMC model. The QMC BEs for the proton hole states are also shown together with
the corresponding experimental values (given within the brackets). 
}
\vspace{0.2cm}
\begin{tabular}{|c|c|c|c|} \hline
 State & BE &$V_v$ &  $V_s$ \\
       &(\footnotesize{MeV})&(\footnotesize{MeV})& (\footnotesize{MeV}) \\
\hline
$^{12}{\!\!\!_{\Xi^-}}$Be$(1s_{1/2})$ &  5.681 & 118.082 & -145.780 \\
$^{28}{\!\!\!_{\Xi^-}}$Mg$(1s_{1/2})$ & 11.376 & 124.674 & -153.881 \\
$^{28}{\!\!\!_{\Xi^-}}$Mg$(1p_{3/2})$ &  5.490 & 167.124 & -206.326 \\
$^{28}{\!\!\!_{\Xi^-}}$Mg$(1p_{1/2})$ &  5.836 & 181.486 & -223.858 \\
$^{12}$C $(1p_{3/2})$                 & 14.329(15.957) & 382.598 & -472.343 \\
$^{28}$Si$(1d_{5/2})$                 & 10.071(11.585) & 378.421 & -467.186 \\
\hline
\end{tabular}
\end{table}
\noindent

The use of bound state spinors calculated within the QMC model provides an 
opportunity to investigate the role of the quark degrees of freedom in the 
cascade hypernuclear production, which has not been done in previous studies
 of this system. Since the cascade hypernuclear 
production involves large momentum transfers (~350 MeV/c - 600 MeV/c) to 
the target nucleus, it is a good case for examining such short distance 
effects. In the QMC model~\cite{gui88}, quarks within the 
non-overlapping nucleon bags (modeled using the MIT bag), interact self 
consistently with isoscalar-scalar ($\sigma$) and isoscalar-vector ($\omega$) 
mesons in the mean field approximation. The explicit treatment of the 
nucleon internal structure represents an important departure from quantum 
hadrodynamics (QHD) model~\cite{ser86}. The self-consistent response of the 
bound quarks to the mean $\sigma$ field leads to a new saturation mechanism 
for nuclear matter~\cite{gui88}. The QMC model has been used to study the 
properties of finite nuclei~\cite{sai96}, the binding of $\omega$, $\eta$, 
$\eta^\prime$ and $D$ nuclei \cite{tsu98a,tsu98b,tsu99} and also the 
effect of the medium on $K^\pm$ and $J/\Psi$ production~\cite{sai07}.

The most recent development of the quark-meson coupling model is the
inclusion of the self-consistent effect of the mean scalar field on the
familiar one-gluon exchange hyperfine interaction that in free space leads
to the $N-\Delta$ and $\Sigma-\Lambda$ mass splitting~\cite{rik07}. With
this~\cite{gui08} the QMC model has been able to explain the properties of
$\Lambda$ hypernuclei for the $s$-states rather well, while the $p$- and
$d$-states tend to underbind.  It also leads to a very natural explanation
of the small spin-orbit force in $\Lambda$-nucleus interaction. In this
exploratory work, the bound $\Xi$ spinors are generated from this version
of the QMC model and are used to calculate the cross sections of the
$^{12}$C($K^-,K^+)^{12}{\!\!\!_{\Xi^-}}$Be  and 
$^{28}$Si($K^-,K^+)^{28}{\!\!\!_{\Xi^-}}$Mg reactions.

To calculate the bound state spinors, we have used the latest version of 
the QMC model. In this version, while the quality of results for $\Lambda$ 
and $\Xi$ is comparable that of the earlier QMC results~\cite{tsu98b}, no 
bound states for the $\Sigma$ states~\cite{gui08} are found. The latter is  
in agreement with the experimental observations. This is facilitated by the 
extra repulsion associated with the increased one-gluon-exchange hyperfine 
interaction in medium. We refer to Ref.~\cite{gui08} for more details of 
this new version of the QMC.

In order to calculate the properties of finite hypernuclei, we construct
a simple, relativistic shell model, with the nucleon core calculated in a
combination of self-consistent scalar and vector mean fields. The
Lagrangian density for a hypernuclear system in the QMC model is
written as a sum of two terms, ${\La}^{HY}_{QMC}$ = ${\La}_{QMC}
+ {\La}^Y_{QMC}$, where
\cite{tsu98a},
\begin{eqnarray}
&&\hspace{-1.4cm}{\La}_{QMC} =  \bar{\psi}_N(\vec{r})
[ i \gamma \cdot \partial - M_N(\sigma) - (\, g_\omega \omega(\vec{r}) \nn \\
&& + g_\rho \frac{\tau^N_3}{2}b(\vec{r})
+ \frac{e}{2} (1+\tau^N_3) A(\vec{r}) \,) \gamma_0 ] \psi_N(\vec{r}) \nn \\
%
&& - \frac{1}{2}[ (\nabla \sigma(\vec{r}))^2 + m_{\sigma}^2
      \sigma(\vec{r})^2 ] \nn \\
&& + \frac{1}{2}[ (\nabla \omega(\vec{r}))^2 + m_{\omega}^2 \omega(\vec{r})^2 ]
\nn \\
&& + \frac{1}{2}[ (\nabla b(\vec{r}))^2 + m_{\rho}^2 b(\vec{r})^2 ]
 +\frac{1}{2} (\nabla A(\vec{r}))^2, \label{Lag2}
\end{eqnarray}
and
\begin{eqnarray} 
&&\hspace{-1.4cm}{\La}^Y_{QMC} = \sum_{Y=\Lambda,\Sigma,\Xi}
\overline{\psi}_Y(\vec{r}) [ i \gamma \cdot \partial - M_Y(\sigma)
- (\, g^Y_\omega \omega(\vec{r})\nn \\
&& + g^Y_\rho I^Y_3 b(\vec{r})
+ e Q_Y A(\vec{r}) \,) \gamma_0 ] \psi_Y(\vec{r}), \qquad \label{Lag3}
\end{eqnarray}
where $\psi_N(\vec{r})$, $\psi_Y(\vec{r})$, $b(\vec{r})$ and $\omega(r)$ 
are, respectively, the nucleon, hyperon, the $\rho$ meson and the $\omega$ 
meson fields, while $m_\sigma$, $m_\omega$ and $m_{\rho}$ are the masses of 
the $\sigma$, $\omega$ and $\rho$ mesons. The $A(r)$ is Coulomb field. 
$g_\omega$ and $g_{\rho}$ are the $\omega$-N and $\rho$-N coupling constants 
which are related to the corresponding (u,d)-quark-$\omega$, $g_\omega^q$, 
and $(u,d)$ quark-$\rho$, $g_\rho^q$, coupling constants as $g_\omega = 3 
g_\omega^q$ and $g_\rho = g_\rho^q$.  $I^Y_3$ and $Q_Y$ are the third 
component of the hyperon isospin operator and its electric charge in units 
of the proton charge, $e$, respectively.

The following set of equations of motion are obtained for the hypernuclear
system from the Lagrangian density Eqs.~(\ref{Lag2})-(\ref{Lag3}):
\begin{eqnarray}
&&\hspace{-1.4cm}[i\gamma \cdot \partial -M_N(\sigma)-
(\, g_\omega \omega(\vec{r}) + g_\rho \frac{\tau^N_3}{2} b(\vec{r}) \nn \\
&&\hspace{-1.4cm} + \frac{e}{2} (1+\tau^N_3)
 A(\vec{r}) \,) \gamma_0 ] \psi_N(\vec{r}) =  0, \label{eqdiracn1}
\end{eqnarray}
\begin{eqnarray}
&&\hspace{-1.4cm}[i\gamma \cdot \partial - M_Y(\sigma)-
(\, g^Y_\omega \omega(\vec{r}) + g_\rho I^Y_3 b(\vec{r}) \nn \\
&&\hspace{-1.4cm} + e Q_Y A(\vec{r}) \,) \gamma_0 ] \psi_Y(\vec{r}) = 0, 
\label{eqdiracy2}
\end{eqnarray}
\begin{eqnarray}
&&\hspace{-1.4cm}(-\nabla^2_r+m^2_\sigma)\sigma(\vec{r})  = \nn \\
&&\hspace{-1.4cm} g_\sigma C_N(\sigma) \rho_s(\vec{r})
 + g^Y_\sigma C_Y(\sigma) \rho^Y_s(\vec{r}), \label{eqsigma}
\end{eqnarray}
\begin{eqnarray}
&&\hspace{-1.4cm}(-\nabla^2_r+m^2_\omega) \omega(\vec{r})  =
g_\omega \rho_B(\vec{r}) + g^Y_\omega
\rho^Y_B(\vec{r}),\label{eqomega}
\end{eqnarray}
\begin{eqnarray}
&&\hspace{-1.4cm}(-\nabla^2_r+m^2_\rho) b(\vec{r})  =
\frac{g_\rho}{2}\rho_3(\vec{r}) + g^Y_\rho I^Y_3 \rho^Y_B(\vec{r}),
 \label{eqrho}
\end{eqnarray}
\begin{eqnarray}
&&\hspace{-1.4cm}(-\nabla^2_r) A(\vec{r})  =
e \rho_p(\vec{r})
+ e Q_Y \rho^Y_B(\vec{r}) ,\label{eqcoulomb}
\end{eqnarray}
where, $\rho_s(\vec{r})$ ($\rho^Y_s(\vec{r})$), $\rho_B(\vec{r})$
($\rho^Y_B(\vec{r})$), $\rho_3(\vec{r})$ and $\rho_p(\vec{r})$ are the
scalar, baryon, third component of isovector, and proton densities at the
position $\vec{r}$ in the hypernucleus~\cite{tsu98a}. On the right hand side
of Eq.~(\ref{eqsigma}), a new and characteristic feature of QMC appears,
arising from the internal structure of the nucleon and hyperon, namely,
$g_\sigma C_N(\sigma)= - \frac{\partial M_N(\sigma)}{\partial \sigma}$
and $g^Y_\sigma C_Y(\sigma)= - \frac{\partial M_Y(\sigma)}{\partial \sigma}$
where $g_\sigma \equiv g_\sigma (\sigma=0)$ and $g^Y_\sigma \equiv
g^Y_\sigma (\sigma=0)$. We use the nucleon and hyperon masses as parameterized
in Ref.~\cite{gui08}. The scalar and vector fields as well as the spinors
for hyperons and nucleons, can be obtained by solving these coupled equations
self-consistently.
\begin{figure}[t]
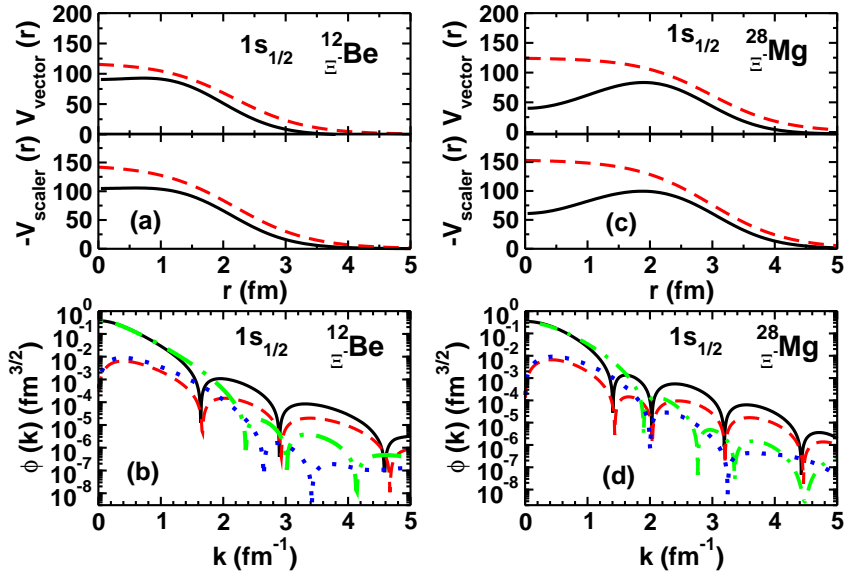

\begin{tabular}{cc}
\centering
\includegraphics[width=.45\textwidth]{Fig2a.eps}\hspace{0.1cm}
\includegraphics[width=.45\textwidth]{Fig2b.eps}
\end{tabular}
\caption{(color online) [(a)] Vector and scalar potential fields for 
$1s_{1/2}$ $\Xi$ state in $^{12}{\!\!\!_{\Xi^-}}$Be. The QMC model and Dirac 
single particle results are shown by solid and dashed lines, respectively. 
[(b)] Moduli of the upper ($|f|$) and lower ($|g|$) components of the 
$1s_{1/2}$ $\Xi$ orbits in $^{12}{\!\!\!_{\Xi^-}}$B hypernucleus in momentum 
space. $|f|$ and $|g|$ of the QMC model are shown by the solid and dashed 
lines, respectively while those of the phenomenological model by the
dashed-dotted and dotted lines, respectively. [(c)] and [d] represent the 
same for $^{28}{\!\!\!_{\Xi^-}}$Mg hypernucleus.}
\label{fig:Fig2}
\end{figure}

In Figs.~2(a) and Figs~2(c), we compare the scalar and vector fields as 
calculated within the QMC model with those of the phenomenological model 
for $1s_{1/2}$ $\Xi^-$ states of $^{12}{\!\!\!_{\Xi^-}}$Be and 
$^{28}{\!\!\!_{\Xi^-}}$Mg hypernuclei, respectively. It may be recalled that 
in the QMC model the scalar and vector fields are generated by the couplings 
of the $\sigma$ and $\omega$ mesons to the quarks. Because of the different 
masses of these mesons and their couplings to the quark fields the scalar 
and vector fields acquire a different radial dependence. In contrast, the two 
fields have the same radial shapes in the phenomenological model. We notice 
that in general, the QMC scalar and vector fields are smaller in magnitude 
than those of the phenomenological model in the entire $r$-region. One 
interesting point to note is that for the heavier hypernucleus, both scalar 
and vector QMC fields have their maxima away from the point $r =0$, in 
contrast to the phenomenological fields. In the mean field models of the 
finite nuclei the proton densities are somewhat pushed out as compared to 
those of the neutron, because of Coulomb repulsion. This causes the $\Xi^-$ 
potential to peak outside the center of the nucleus. This is a consequence 
of the self consistent procedure. In the case of a chargeless hyperon 
(e.g. $\Lambda$) such effects are not observed.

In Figs.~2(b) and 2(d) the moduli of the upper and lower components of 
$1s_{1/2}$ $\Xi^-$ momentum space QMC (solid and dashed line) and 
phenomenological (dashed-dotted and dotted) spinors are shown for the 
$^{12}{\!\!\!_{\Xi^-}}$Be and $^{28}{\!\!\!_{\Xi^-}}$Mg hypernuclei, 
respectively. It is seen that the spinors of the two models are similar to
each other for momenta ($k$) up to 2.0 fm$^{-1}$.  Beyond this region, 
however, they start having differences. The position of minima in the 
phenomenological model spinors is shifted to higher values of $k$ and 
their magnitudes are smaller than those of the QMC model. It should however,
be remarked here that the structure of the minima reflects the size of the
system. An improved search for the depths of the WS potentials in the 
phenomenological model as discussed, above might remove the differences seen 
between the spinors of the two models. We further note that only for $k$ 
values below 1.5 fm$^{-1}$, are the magnitudes of the lower components, 
$|g(k)|$, substantially smaller than those of the upper components. In the 
region of $k$ pertinent to the cascade hypernuclear production, $|g(k)|$ 
may not be negligible. Thus the relativistic effects resulting from the 
small component of bound states spinors could be large for the hypernuclear 
production reactions on nuclei (see also the discussions presented in 
Ref.~\cite{ben89}).

\subsection{Cross Sections for Hypernuclear Production}

In order to calculate the amplitudes (and hence the cross sections)
of the hypernuclear production reaction (see Fig.~ 1b), one requires the 
effective Lagrangians at the meson-baryon-resonance vertices and the 
corresponding coupling constants, and also the propagators for various 
resonances. After having established these quantities the amplitudes of 
the graphs of the type shown in Fig.~1 can be written by following the well 
known Feynman diagrams and can be computed numerically.
 
The effective Lagrangians for the resonance-kaon-baryon vertices for 
spin-$\frac{1}{2}$ and spin-$\frac{3}{2}$ resonances are taken as
\begin{eqnarray}
{\La}_{KBR_{1/2}} & = & -g_{KBR_{1/2}} \bar{\psi}_{R_{1/2}}
                            [\chi\,{i\Gamma}\, {\varphi_K}+
      \frac{(1-\chi)}{M}\,\Gamma\, \gamma_\mu\,(\partial^\mu \varphi_K)]
                            {\psi}_B,\\
{\La}_{KBR_{3/2}} & = & \frac{g_{KBR_{3/2}}}{m_K} \bar{\psi}_{R_{3/2}}^\mu 
        \partial_\mu \phi_{K} \psi_{B} + \text{h.~c.},
\end{eqnarray}
with $M \,=\,(m_R \,\pm\,m_B)$, where the upper sign corresponds to an
even-parity and the lower sign to an odd-parity resonance, and B 
represents either a nucleon or a $\Xi$ hyperon. The operator $\Gamma$ is 
$\gamma_5$ (1) for an even- (odd-) parity resonance. The parameter $\chi$ 
controls the admixture of pseudoscalar and pseudovector components. The 
value of this parameter is taken to be 0.5 for the $\Lambda^*$ and 
$\Sigma^*$ states, but zero for $\Lambda$ and $\Sigma$ states, implying 
pure pseudovector couplings for the corresponding vertices in 
agreement with Refs.~\cite{shy08,shy99}. It may be noted that the  
Lagrangian for spin-$\frac{3}{2}$ as given by Eq.~(10) corresponds to that 
of a pure Rarita-Swinger form which has been used in all previous 
calculations of the hypernuclear production reactions within a similar 
effective Lagrangian model~\cite{shy04,shy08,shy09}. 

Similar to Ref.~\cite{shy11}, we have used the following form factor at 
various vertices, 
\begin{eqnarray}
  F_m(s)=\frac{\lambda^4}{\lambda^4+(s-m^2)^2},
\end{eqnarray}
where $m$ is the mass of the propagating particle and $\lambda$ is the 
cutoff parameter, which is taken to be 1.2 GeV everywhere which is the 
same as that used in Ref.~\cite{shy11}.

The parameters of the resonance vertices were fixed in Ref.~\cite{shy11} 
by describing the total cross section data on elementary reactions
$p(K^-,K^+)\Xi^-$ and $p(K^-,K^0)\Xi^0$, where the form of the 
spin-$\frac{3}{2}$ interaction vertex was somewhat different form that 
given Eq.~(10). In this paper, therefore, we recalculate the cross sections 
of the elementary reaction using the spin-$\frac{3}{2}$ Lagrangian given 
by Eq.~(10). Apart from the total cross sections, we also describe the 
differential cross sections of the $p(K^-,K^+)\Xi^-$ reaction which was not
done in Ref.~\cite{shy11}. The values of the vertex parameters were taken 
to be the same as those determined in Ref.~\cite{shy11} except for the 
vertices involving the $\Sigma (1385)$ resonance, where the coupling
constants (CCs) have been slightly increased in order to better describe  
the differential cross section data (see Table 2). 
\begin{table}[t]
\begin{center}
\caption{\label{table2} $\Lambda$ and $\Sigma$ resonance intermediate states 
included in the calculations.}
\vspace{0.2cm}
\begin{tabular}{|c|c|c|c|c|c|}
\hline
Intermediate state & $L_{I 2J}$ &$M$      &${Width}$    &$g_{KRN}$ 
&$g_{KR\Xi}$\\
 ($R$)                  &          &(\footnotesize{GeV})&(\footnotesize{GeV})&
  &
\\
\hline
$\Lambda$ &                   &1.116    &0.0         &-16.750  & 10.132  \\
$\Sigma $ &                   &1.189    &0.0         &  5.580  &-13.50   \\
$\Sigma (1385)$ &$ P_{13}$    &1.383    &0.036       &-8.22    & -8.220  \\ 
$\Lambda(1405)$ &$ S_{01}$    &1.406    &0.050       &1.585    & -0.956  \\ 
$\Lambda(1520)$ &$ D_{03}$    &1.520    &0.016       &27.46    &-16.610  \\ 
$\Lambda(1810)$ &$ P_{01}$    &1.810    &0.150       &2.800    &  2.800  \\ 
\hline
\end{tabular}
\end{center}
\end{table}

The two interaction vertices of Fig.~\ref{fig:Fig1} are connected by a 
resonance propagator. For the spin-$1/2$ and spin-$3/2$ resonances the 
propagators are given by
\begin{equation}\label{eq:propspin12}
\mathcal{D}_{R_{1/2}} = \frac{\im (\gamma_\mu p^\mu + m_{R_{1/2}})}
{p^2 - (m_{R_{1/2}} - \im\Gamma_{R_{1/2}}/2)^2},
\end{equation}
and
\begin{equation}\label{eq:propspin32}
\mathcal{D}_{R_{3/2}}^{\mu\nu} = - 
 \frac{\im (\gamma_\lambda p^\lambda + m_{R_{3/2}})}
{p^2 - (m_{R_{3/2}} - \im\Gamma_{R_{3/2}}/2)^2} P^{\mu\nu} \;,
\end{equation}
respectively. In Eq.~\eqref{eq:propspin32} we have defined
\begin{equation}\label{eq:prop32proj}
P^{\mu\nu} =
 g^{\mu\nu} - \frac{1}{3} \gamma^\mu \gamma^\nu - \frac{2}{3m_{R_{3/2}}^2} 
p^\mu p^\nu + \frac{1}{3m_{R_{3/2}}} \left( p^\mu \gamma^\nu - p^\nu \gamma^\mu \right)
\;.
\end{equation}

In Eqs.~\eqref{eq:propspin12} and~\eqref{eq:propspin32}, $\Gamma_{R_{1/2}}$ 
and $\Gamma_{R_{3/2}}$ define the total widths of the corresponding 
resonances. We have ignored any medium modification of the resonance widths
while calculating the amplitudes of the hypernuclear production as 
information about them is scarce and uncertain. 

In the next section we describe the results of our calculations for the 
$(K^-,K^+)$ reaction on both proton and nuclear targets.

\section{Results and Discussions}

In Figs.~3a, we show comparisons of our calculations with the data for the 
total cross section of the $p(K^-, K^+)\Xi^-$ reaction for $K^-$ beam 
momenta ($p_{K^-}$) below 3.5 GeV/c, because the resonance picture is not 
suitable at momenta higher than this. It is clear that our model is able 
to describe well the beam momentum dependence of the total cross section 
data of the elementary reactions within statistical errors.  The arrow 
in Fig.~3a shows the position of the threshold beam momentum for this 
reaction which is about 1.0 GeV. The measured total cross section peaks in 
the region of 1.35-1.4 GeV/c which is well described by our model. We further
note that the cross sections for  $p_{K^-} < 2.0$ GeV/c are dominated by 
the $s$-channel contributions.
 
In Fig. 3b we  compare our calculations with the differential cross section 
data of the $p(K^-, K^+)\Xi^-$ reaction for $p_{K^-}$ values of 1.7 GeV/c 
and 2.1 GeV/c. These data were read from the corresponding figures given in 
Ref.~\cite{dau69}. Both calculated and experimental differential cross 
sections are normalized to the same total cross section. We see that our 
calculations describe the general trends of the angular distribution data
well in the entire angular region for both the beam momenta. Nevertheless, 
a slight overestimate of the data is noted at the forward angles. There 
is a need to remeasure these differential cross sections at the JPARC 
facility to confirm and refine the old bubble chamber data of 
Ref.~\cite{dau69}.  
\begin{figure}[t]
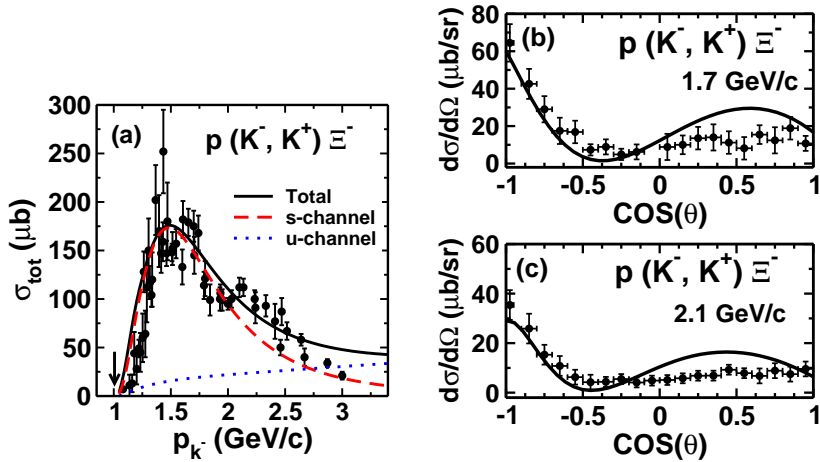

\begin{tabular}{cc}
\centering
\includegraphics[width=.42\textwidth]{Fig3a.eps}\hspace{0.50cm}
\includegraphics[width=.42\textwidth]{Fig3b.eps}
\end{tabular}
\caption{(color online)
(a) Comparison of the calculated total cross section for the 
$p(K^-, K^+)\Xi^-$ reaction as a function of incident $K^-$ 
momentum with the corresponding experimental data. Also shown are the
individual contributions of $s$- and $u$-channel diagrams to the total
cross section. The arrow indicates the position of the threshold for this
reaction. (b) and (c) Differential cross sections for the same reaction for
$K^-$ beam momenta of 1.7 GeV/c and 2.1 GeV/c, respectively. 
}
\label{fig:Fig3}
\end{figure}

The beam momentum dependence of the $0^\circ$ differential cross section 
($d\sigma/d\Omega)_{0^\circ}$ for the $p(K^-,K^+)\Xi^-$ reaction is an 
interesting quantity because it enters explicitly into the expression for 
the cross sections of the $(K^-,K^+)$ reaction on nuclei (leading to the 
production of $\Xi$ hypernuclei) in the kind of model used in Ref.
\cite{gal83}.  Hence, the beam energy dependence of the zero 
angle differential cross section of the hypernuclear production directly 
follows that of [($d\sigma/d\Omega)_{0^\circ}$]. In Fig.~4, we show the beam 
momentum dependence of this quantity (using the same normalization as those 
in Figs. 3b and 3c). We see that [($d\sigma/d\Omega)_{0^\circ}$] peaks in the 
same region of $p_{K^-}$ as the total cross section shown in Fig.~3a. On 
the other hand, the situation regarding the momentum dependence of the 
available experimental data on [($d\sigma/d\Omega)_{0^\circ}$] is quite 
uncertain.  The existing data reported in Refs.~\cite{ber66,bur68,dau69} 
differ considerably from each other. This may be due to normalization problems 
between different experiments or may be arising from the large errors in 
the Legendre coefficients. In fact only Ref.~\cite{dau69} shows 
the data explicitly for four values of $p_{K^-}$ between 1.7 GeV/c to 2.64 
GeV/c, together with the coefficients of the Legendre polynomial fits to 
the data. The other two references give only the coefficients of such a 
fit that have large correlated errors. The cross sections of Ref.
\cite{ber66} could have maxima at both 1.4 GeV/c and 1.7 GeV/c within 
the statistical errors. The data of Ref.~\cite{bur68} have a peak at 1.74
GeV/c but the coefficients of the Legendre polynomial fits given there may 
have misprints - at one beam momentum they even give negative cross section.
Therefore, the position of the peak in the experimental zero degree 
differential cross section of the $p(K^-,K^+)\Xi^-$ is uncertain. A proper 
measurement of this quantity at the JPARC facility would be very welcome in 
order to remove this anomaly. 

\begin{figure}[t]
\centering
\includegraphics[width=.50\textwidth]{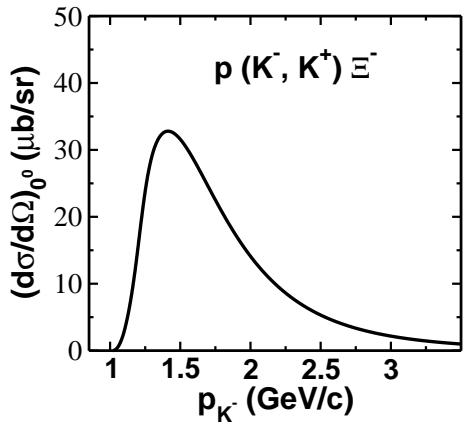}
\caption{(color online)
The zero degree differential cross section of the $p(K^-, K^+)\Xi^-$ 
reaction calculated as a function of beam momentum. 
}
\label{fig:Fig4}
\end{figure}

In calculations of the hypernuclear production reactions, we have employed 
pure single-particle-single-hole $(\Xi p^{-1})$ wave functions to describe 
the nuclear structure part, ignoring any configuration mixing effects. The 
nuclear structure part is treated exactly in the same way as described in 
Ref.~\cite{shy08}. The amplitude involves the momentum space four component
(spin space) spinors ($\psi$) which represent the wave functions of the bound
states of nucleon and hyperon. For the proton hole and $\Xi^-$ states, 
spinors generated within the QMC and the phenomenological models were used in 
the respective calculations. We have used a plane wave approximation to 
describe the relative motion of kaons in the incoming and outgoing channels. 
However, the distortion effects are partially accounted for by introducing 
reduction factors to the cross sections as described in Ref.~\cite{ike94}. 
Since our calculations are carried out all along in momentum space, they 
include all the nonlocalities in the production amplitudes that arise from 
the resonance propagators.
\begin{figure}
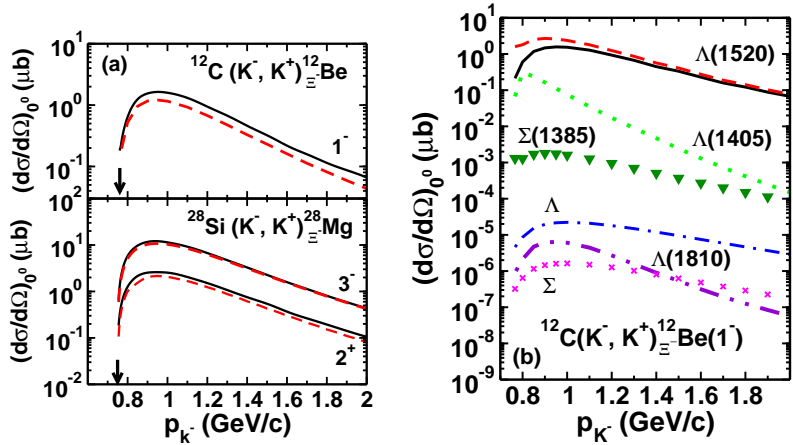

\begin{tabular}{cc}
\centering
\includegraphics[width=.40\textwidth]{Fig5a.eps}\hspace{0.38cm}
\includegraphics[width=.42\textwidth]{Fig5b.eps}
\end{tabular}
\caption{(color online)
(a) Calculated differential cross section at $0^\circ$ as a function of 
$K^-$ beam momentum for the $^{12}$C$(K^-,K^+)^{12}{\!\!\!_{\Xi^-}}$Be
and $^{28}$Si$(K^-,K^+)^{28}{\!\!\!_{\Xi^-}}$Mg reactions. The spin-parity 
of the final hypernuclear states are indicated on each curve. The 
particle-hole configurations of the $^{12}{\!\!\!_{\Xi^-}}$Be($1^-$) and 
the $^{28}{\!\!\!_{\Xi^-}}$Mg($2^+$) hypernuclear states are 
$[(1p_{3/2})_p^{-1},(1s_{1/2})_{\Xi^-}]1^{-}$, and 
$[(1d_{5/2})_p^{-1},(1s_{1/2})_{\Xi^-}]2^{+}$, respectively. For the 
$^{28}{\!\!\!_{\Xi^-}}$Mg($3^-$) state the results shown are the sum of 
the cross section corresponding to both 
$[(1d_{5/2})_p^{-1},(1p_{3/2})_{\Xi^-}]3^{-}$ and
$[(1d_{5/2})_p^{-1},(1p_{1/2})_{\Xi^-}]3^{-}$ configurations. 
The solid and dashed lines represent the results obtained with the QMC and 
the phenomenological $\Xi^-$ bound state spinors. Arrows show the threshold 
of the two reactions. (b) Contributions of individual resonance intermediate 
states as indicated near each line, to the zero angle differential cross 
section as function of $K^-$ beam momentum for the 
$^{12}$C$(K^-,K^+)^{12}{\!\!\!_{\Xi^-}}$Be(1$^-$) reaction. Their coherent
sum is shown by the solid line.  
}
\label{fig:Fig5}
\end{figure}

We have chosen the reactions $^{12}$C$(K^-,K^+)^{12}{\!\!\!_{\Xi^-}}$Be
and $^{28}$Si$(K^-,K^+)^{28}{\!\!\!_{\Xi^-}}$Mg for the first
application of our model. The reaction on the $^{12}$C target is billed
as the "day one" experiment at the JPARC facility. The thresholds for these
reactions are about 0.761 GeV/c and 0.750 GeV/c, respectively and the
momentum transfers involved at 0$^\circ$, vary between 1.8 - 2.9 
fm$^{-1}$. The initial states in both the cases are doubly closed systems. 
The QMC model predicts only one bound state for the $^{12}{\!\!\!_{\Xi^-}}$Be 
hypernucleus with the $\Xi^-$ hyperon being in a 1$s_{1/2}$ state with a 
binding energy as shown in Table 1. For the $^{28}{\!\!\!_{\Xi^-}}$Mg 
case however, three distinct bound $\Xi^-$ states with configurations 
1$s_{1/2}$, 1$p_{3/2}$ and 1$p_{1/2}$ have been predicted. The binding 
energies of these states are shown in Table 1. It is evident that for 
this nucleus 1$p_{3/2}$ and 1$p_{1/2}$ states are almost degenerate. This 
reflects the fact that the $\Xi$-nucleus spin-orbit potential is weak. This 
is due to the fact that since the corresponding total potential depth is 
small, the gradient of this potential that contributes to the spin-orbit 
force is also small.

In case of the $^{12}$C target, the $\Xi^-$ hyperon in a 1$s_{1/2}$ state  
can populate 1$^-$ and 2$^-$ states of the hypernucleus corresponding to 
the particle-hole configuration $[(1p_{3/2})^{-1}_p,(1s_{1/2})_{\Xi^-}$]. 
The states populated for the $^{28}{\!\!\!_{\Xi^-}}$Mg hypernucleus are  
[2$^+$, 3$^+$], [1$^-$, 2$^-$, 3$^-$, 4$^-$], and [$2^-$, $3^-$] 
corresponding to the configurations $[(1d_{5/2})^{-1}_p, (1s_{1/2})_{\Xi^-}$],
$[(1d_{5/2})^{-1}_p,(1p_{3/2})_{\Xi^-}$], and $[(1d_{5/2})^{-1}_p,
(1p_{1/2})_{\Xi^-}$], respectively. In Fig.~5, we have shown results for 
populating the hypernuclear state with maximum spin of natural parity for 
each configuration. The unnatural parity states are very weakly excited due 
to the vanishingly small spin-flip amplitudes for this reaction (see, e.g., 
Ref.~\cite{gal83} for an extensive discussion on this point). However, for 
the $^{28}{\!\!\!_{\Xi^-}}$Mg($3^-$) case, the results shown are the sum of 
the cross sections obtained with both $[(1d_{5/2})^{-1}_p,(1p_{3/2})_{\Xi^-}$]
and $[(1d_{5/2})^{-1}_p,(1p_{1/2})_{\Xi^-}$] particle-hole configurations. 
The latter contributes substantially (up to about 75$\%$ within our model) to 
the excitation of this state. 

In Fig.~5a, the 0$^\circ$ differential cross sections are shown as a function
of the beam momentum that are obtained by using $\Xi^-$ bound state spinors 
calculated within the QMC as well as the phenomenological model for the 
reactions $^{12}$C$(K^-,K^+)^{12}{\!\!\!_{\Xi^-}}$Be and 
$^{28}$Si$(K^-,K^+)^{28}{\!\!\!_{\Xi^-}}$Mg. The configurations of the 
final hypernuclear states are as described in the figure caption. 
In the calculations of our reaction amplitudes, the relative motions
of $K^-$ and $K^+$ mesons in the initial and final channels, respectively 
are described by plane waves. The distortion effects, which primarily 
describe the absorption of the incoming $K^-$, are however, included by 
introducing factors that reduce the magnitudes of the cross sections. These
factors are taken to be 2.8 and 5.0 for $^{12}$C and $^{28}$Si targets, 
respectively as suggested in Ref.~\cite{ike94}. This necessarily assumes
that shapes of the angular distributions are not affected by the distortion
effects. This aspect will be further investigated in a future study. 

We see that the QMC model cross sections are larger than those obtained by 
using the phenomenological model by about 10-15$\%$ in all the cases. This 
reflects the fact that in the region of momentum transfers relevant to these 
reactions both the upper and the lower components of the QMC spinors are 
higher in magnitude than the corresponding phenomenological ones. 

An important observation in Fig. 5a is that for both the hypernuclear 
production reactions, the cross sections peak at $p_{K^-}$ around 1.0 GeV/c, 
which is about 0.25-0.26 GeV/c above the production thresholds of the two 
reactions. Interestingly, it is not too different from the case of the 
elementary $\Xi^-$ production reaction where the peaks of the total 
cross section as well as the zero degree differential cross section occur 
at about 0.35-0.40 GeV/c above the corresponding production threshold 
(see Figa.~3a and 4). Furthermore, the magnitudes of the cross sections 
near the peak position are in excess of 1 $\mu b$. It is important in this 
context to note that the magnitude of our cross section for a $^{12}$C target
at a beam momentum of 1.6 GeV/c is similar to that  obtained in Ref.
\cite{ike94} within an impulse approximation model. Moreover, our cross 
sections at 1.8 GeV/c also are very close those of Ref.~\cite{gal83} for 
both the targets. However, we fail to corroborate the results of 
Ref.~\cite{gal83} where cross sections were shown to peak for $p_{K^-}$ 
around 1.8 GeV/c. It is quite probable that the distortion effects are 
dependent on the beam momenta and may be relatively stronger at lower 
values of $p_{K^-}$. Nevertheless, this is unlikely to lead to such a large 
shift in the peak position. In any case, this effect was not considered in 
Ref.~\cite{gal83} also. There may be a need to re-examine the beam momentum 
dependence of the zero degree differential cross section  in order to 
understand this different.

In Fig~5b, we note that the contribution from the $\Lambda(1520)$
intermediate state dominates the total cross sections over the entire 
regime of $p_{K^-}$ values. This is similar to that noted in the case 
of the elementary $\Xi^-$ production reaction. The $\Lambda(1405)$, and 
$\Sigma(1385)$ states make noticeable contributions only for $p_{K^-}$ very 
close to the production threshold. Other resonances contribute very weakly. 
Of course, our results are quite dependent on the CCs of various vertices, 
which are somewhat uncertain. Nevertheless, the respective cross sections 
shown in this figure are robust. First of all these CCs provide a 
good description of the total as well a differential cross sections of the 
elementary $\Xi^-$ production reaction. Secondly there is very little scope 
for increasing further the individual contributions of the $\Lambda$ and 
$\Sigma$ intermediate states, because the CCs of the corresponding vertices
used by us are already larger than their upper limits suggested in the 
literature. Furthermore,  the contributions of other resonances are too weak 
and even have the wrong $p_{K^-}$ dependence. Therefore, the final results 
are unlikely to be affected too much by the known uncertainties in the 
corresponding CCs.

\section{Summary and Conclusions}

In summary, in this paper the cascade hypernuclear production reactions  
$^{12}$C($K^-,K^+)^{12}{\!\!\!_{\Xi^-}}$Be, and 
$^{28}$Si($K^-,K^+)^{28}{\!\!\!_{\Xi^-}}$Mg have been studied within an 
effective Lagrangian model, using the proton hole and $\Xi^-$ bound state 
spinors derived from the latest quark-meson coupling model. This is for 
the first time that the quark degrees of freedom have been explicitly 
invoked in the description of such reactions. We have considered the 
excitation of altogether six $\Lambda$ and $\Sigma$ hyperon resonance 
intermediate states in the initial collision of the $K^-$ meson with a 
target proton. These states subsequently propagate and decay into a 
$\Xi^-$ hyperon and a $K^+$ meson. The hyperon gets captured in one of 
the nuclear orbits, while the meson goes out. We constrain the coupling 
constants at the resonance vertices by describing both the total and the 
differential cross sections of the elementary $p(K^+,K^-)\Xi^-$ reaction 
within a similar model. 

We have also performed calculations with the spinors obtained by solving 
the Dirac equation with vector and scalar potential fields having 
Woods-Saxon shapes (the phenomenological model). Their depths are 
fitted to the binding energies of the respective states (QMC model 
values for the $\Xi^-$ particle states and experimental values for the proton
hole states) for a given set of geometry parameters which are taken to 
be the same for the two fields. While for $^{12}{\!\!\!_{\Xi^-}}$Be 
hypernucleus the shapes of the QMC fields are similar to those of the 
phenomenological model, the two differ considerably in the case of 
$^{28}{\!\!\!_{\Xi^-}}$Mg. For the cases studied in this paper, the 
hypernuclear production cross sections calculated with the QMC $\Xi^-$ 
spinors are found to differ only slightly from those obtained within the 
phenomenological (the former being about 10-15$\%$ higher in  magnitude than
the later). The distortion effects are included by introducing
reduction factors to the cross sections taken from the previous studies of
this reaction.  

The zero degree differential cross sections for the $\Xi^-$ hypernuclear 
production reactions on the two targets considered here, have peaks around 
the beam momentum of 1.0 GeV/c within both the QMC and the phenomenological 
models. This peak momentum is above the corresponding production threshold by 
almost the same amount as the position of the maximum in the elementary 
total as well as zero degree differential cross sections lies away from its 
respective threshold. The peak cross sections are in excess of 1 $\mu b$. 
Furthermore, the total hypernuclear production cross sections are dominated 
by the contributions from the $\Lambda(1520)$ ($D_{03}$) resonance 
intermediate state which is similar to the case of the elementary $\Xi^-$ 
production reaction. Other resonances make noticeable contributions only at 
beam momenta close to the production threshold of the reaction. It is 
desirable to perform measurements for the differential cross sections of the 
elementary $\Xi^-$ production reaction in a wide beam momentum range. 

This work was supported by the University of Adelaide and the Australian 
Research Council through grant FL0992247(AWT).

\end{document}